*To be submitted to Appl. Phys. Lett.*

# Domain Dynamics in Piezoresponse Force Microscopy: Quantitative Deconvolution and Hysteresis Loop Fine Structure

Igor Bdikin and Andrei Kholkin

Department of Ceramics and Glass Engineering, CICECO, University of Aveiro,

3810-193 Aveiro, Portugal, e-mail: kholkin@ua.pt

Anna N. Morozovska* and Sergei V. Svechnikov

V. Lashkaryov Institute of Semiconductor Physics, National Academy of Science of Ukraine,

41, Prospect, Nauki, 03028 Kiev, Ukraine

Seung-Hyun Kim

INOSTEK Inc., Gyeonggi Technopark, Ansan, Gyeonggi 426-901, S. Korea

Sergei V. Kalinin†

The Center for Nanoscale Materials Sciences and Materials Science and Technology

Division, Oak Ridge National Laboratory, Oak Ridge, TN 37831, U.S.A.

* Corresponding author, morozo@i.com.ua

† Corresponding author, sergei2@ornl.gov

Domain dynamics in the Piezoresponse Force Spectroscopy (PFS) experiment is studied using the combination of local hysteresis loop acquisition with simultaneous domain imaging. The analytical theory for PFS signal from domain of arbitrary cross-section is developed and used for the analysis of experimental data on $Pb(Zr,Ti)O_3$ polycrystalline films. The results suggest formation of oblate domain at early stage of the domain nucleation and growth, consistent with efficient screening of depolarization field within the material. The fine structure of the hysteresis loop is shown to be related to the observed jumps in the domain geometry during domain wall propagation (nanoscale Barkhausen jumps), indicative of strong domain-defect interactions.

Local polarization switching in ferroelectric materials by a biased Piezoresponse Force Microscopy (PFM) tip has emerged as a perspective approach for ultra high density data storage,[1,2] ferroelectric lithography,[3] and nanostructure fabrication.[4,5] Subsequent imaging of the switched domain provides direct insight into domain growth mechanism[6] and relaxation kinetics[7], from which the effect of macroscopic defects,[8] microscopic disorder,[9] and surface state on domain wall dynamics and pinning can be established. The primary limitation of this approach is the (a) uncontrollable effect of surface charge migration[10,11] inevitable in ambient conditions on observed growth and relaxation kinetics, and (b) extremely large times (~10s – hours) required for studies even at a single location. Single-point Piezoresponse Force Spectroscopy (PFS) allows probing the domain growth process by detecting the changes in local electromechanical response due to nucleation and growth of domain below the tip. Increase of acquisition speed to ~0.1 – 1 s/loop[12] enabled Switching-Spectroscopy PFM, in which PFS loops are collected at each point in the image to yield 2D maps of parameters such as imprint, nucleation biases, or work of switching.

The development of PFS and SSPFM necessitated the quantitative interpretation of the spectroscopic data, establishing the relationship between the electromechanical response and the geometric parameters of the domain formed below the tip. Recently, the theoretical framework for the interpretation of the PFS data for cylindrical domain geometry was established in Ref.[13]. Here, we report the results of the combined spectroscopic-imaging experiments and correlate the changes in domain geometry and the loop shape.

The PFM spectroscopy and domain imaging at each step of spectroscopic process were preformed as described elsewhere.[14] The films of $Pb(Zr_{0.3}Ti_{0.7})O_3$ (PZT) used in this study were produced by sol-gel method and commercialized by INOSTEK [15]. The thickness of the films was about 4 μm and grain size varied in the range 100-500 nm. The films had columnar structure with random orientation of the grains. Sufficiently big grains of the size about 300-400 nm with strong initial piezoelectric contrast were used for the local measurements. The representative local PFS hysteresis loop, corresponding evolution of domain sizes, and *in-situ* domain images are shown in Fig.1. The data were taken with poling pulse duration 1 s and imaging of the resulting domain structure immediately after poling. Note that a sharp change in response in PFS hysteresis loop at 10-15 V corresponds to domain nucleation. The subsequent lateral growth of the domain corresponds to the increase in PFM signal, which finally saturates at about 30 V.

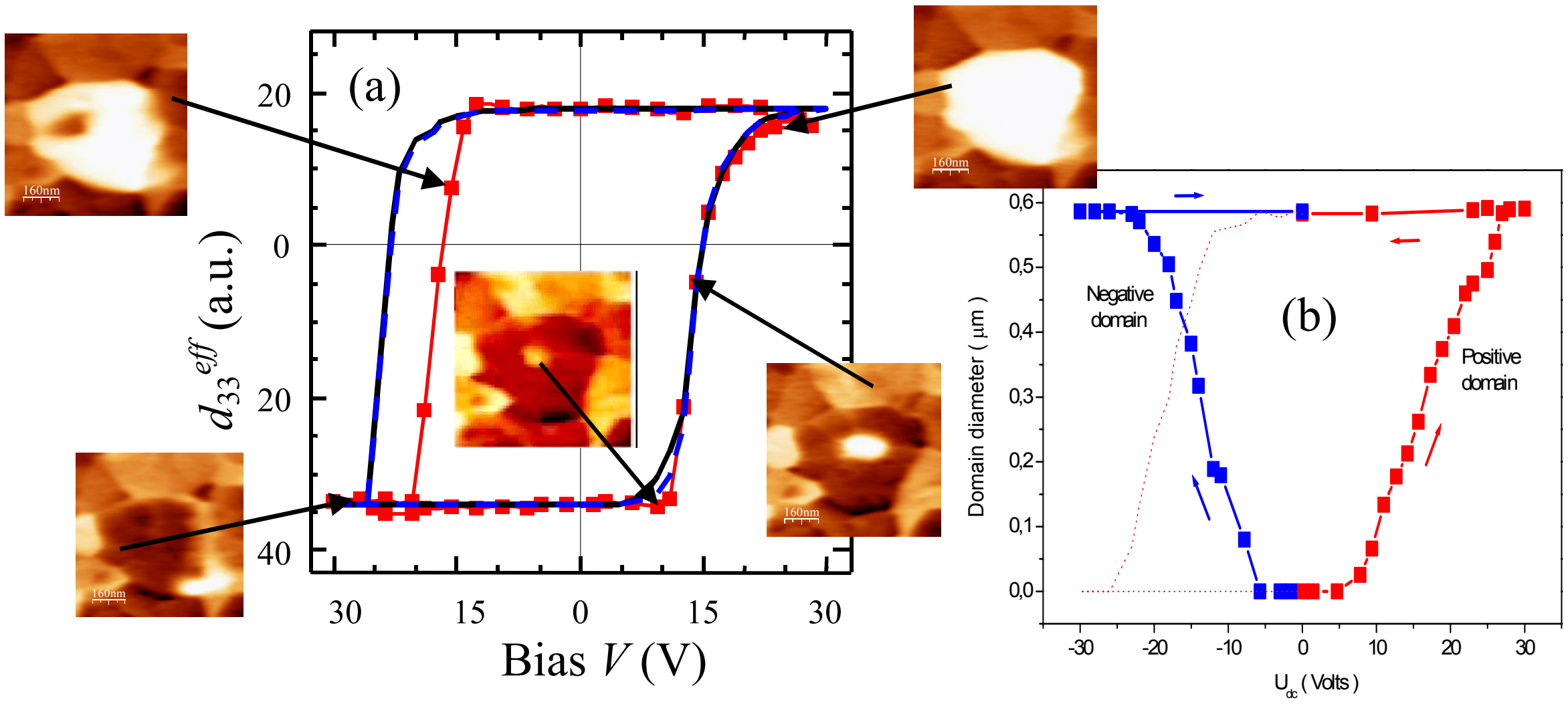

**Fig. 1.** (a) Effective piezoresponse $d_{33}^{eff}$ via applied bias $V$: filled squires is experimental data for examined $Pb(Zr_{0.3}Ti_{0.7})O_3$ 4μm film[16]; solid and dashed curves are theoretical fitting based on Eqs.(1-2) and Eqs.(A.4), respectively. Domain radius $r_s$ for both curves was determined from PFM images and shown in part (b). Solid curve in Fig. 1(a) corresponds to the size of nascent domain (dark and bright), while dashed curve is calculated for the difference between the fully switched area for bright domain and nascent dark domain.

To correlate local electromechanical response measured in a PFS experiment to the size of ferroelectric domain formed below the tip, we utilized the decoupling approximation[17,18] combined with Pade approximation method. Here, we develop the generalized analytical relationship between the PFM signal and effective domain size $R_S(V)$, as:

$$d_{33}^{eff}\left(R_S(V)\right) \approx \left(\frac{3}{4}d_{33} + (1+4\nu)\frac{d_{31}}{4}\right)\frac{\pi d - 8R_S(V)}{\pi d + 8R_S(V)} + \frac{d_{15}}{4}\frac{3\pi d - 8R_S(V)}{3\pi d + 8R_S(V)}. \quad (1)$$

Equation (1) is valid for dielectrically isotropic materials with $\gamma \approx 1$[19], $d_{nm}$ are strain piezoelectric coefficients, and $\nu \approx 0.35$ is the Poisson ratio. The bias-dependent effective domain size $R_S(V)$ was derived as

$$R_S(V) \approx \frac{(\pi/6) l r_S}{\sqrt{r_S^2 + l^2 (\pi/6)^2}}, \quad (2)$$

where $l(V)$ is the effective domain length, radius $r_S(V) \approx \frac{1}{2\pi}\int_0^{2\pi} d\alpha \cdot r(\alpha)$, $\alpha$ is the angle, as defined in Fig. 2(a).

Below we compare the hysteresis loop measurements with the *in-situ* imaging studies. For domains of irregular shape, effective radius, $r_s$, was determined experimentally as a square root of domain cross-section determined from PFM images. Additionally, we introduce the domain center shift with respect to the probe apex axis, $a(V)$, that can be significant at early stages of domain growth process due to the presence of local defects and inhomogeneities. Here, we regard $a(V)$ as a fitting parameter. For high aspect ratio ($r_s/l \ll 1$) tip-induced domains $R_s(V) \approx r_s(V)$ and $a(V) \approx 0$, so the radius $r_s$ can be directly extracted from effective piezoresponse $d_{33}^{eff}(V)$ deconvolution by means of Eqs.(1-2). For small shifts $a \ll r_s$, it is possible to estimate domain length using both effective radius $r_s(V)$ extracted from PFM images and $R_s(V)$ extracted from piezoresponse data $d_{33}^{eff}(V)$ as following:

$$l(V) \approx \frac{6}{\pi}\frac{R_S r_S}{\sqrt{r_S^2 - R_S^2}}. \qquad (3)$$

Domain parameters determination for PZT films is shown in Fig. 2. The accuracy of the proposed deconvolution method and results presented in Fig. 2 can be estimated from Fig.1(a). Note that the solid and dashed curves in Fig.1(a) do not coincide with experimental symbols at negative bias, indicating that the contribution of "small" dark domain into piezoresponse was underestimated (actually we have *in-situ* PFM data for $r_s$ of "bright" domain only). At the same time, dotted and solid curve coincide almost everywhere, indicative of relatively minor contribution of domain center displacement, $a(V)$, to measured signal.

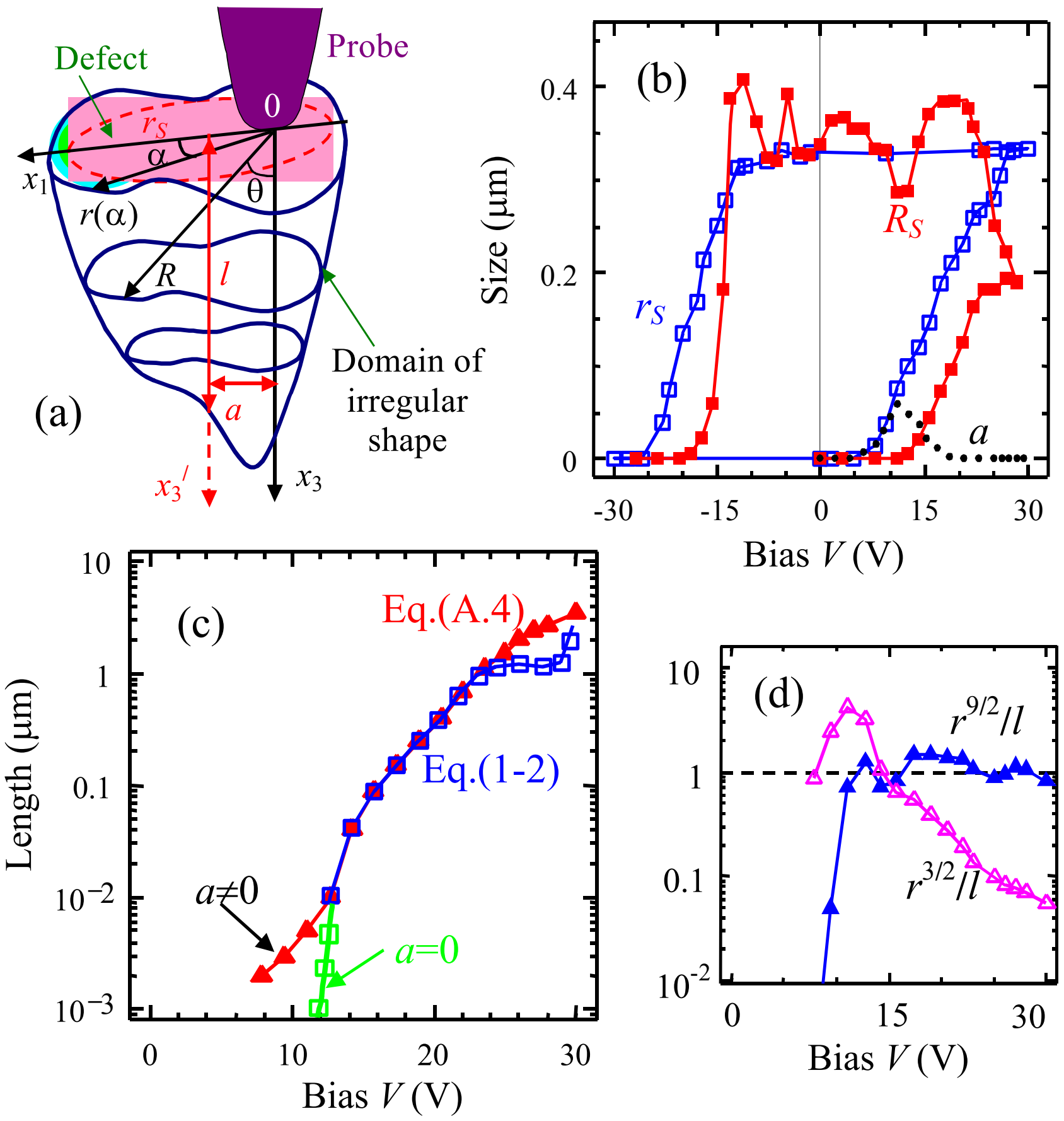

**Fig.2**. (a) Problem geometry. (b) Effective domain size via applied bias: filled squires are $R_S$ values deconvoluted from $d_{33}^{eff}$ using Eq.(1); empty squires are average radius $r_S$ values obtained from *in-situ* PFM data shown in Fig. 1b; dotted curve is domain shift $a$ deconvoluted from $d_{33}^{eff}$ using *in-situ* $r_S$ values in Eqs.(A.4). (c) Effective domain length via applied bias: filled triangles were deconvoluted from $d_{33}^{eff}$ using *in-situ* $r_S$ values in Eqs.(A.4) and a≠0; empty squares are deconvoluted from $d_{33}^{eff}$ using *in-situ* $r_S$ values in Eqs.(1-2) and a=0. (d) The ratios $r_S^{3/2}/l$ (invariant for prolate domains in Molotskii model[20]) and $r_S^{9/2}/l$ (almost constant in our case) calculated from $r_S$ values in Eqs.(A.4) (labels near the curves, arbitrary units). For $Pb(Zr_{0.3}Ti_{0.7})O_3$ constants $d_{31}$=-11.4, $d_{33}$=61.2 pm/V[15] $\gamma \approx 1$ and $d$=40nm; vertical offset caused by electrostatic forces is -6.0 a.u. The deconvolution results are independent of $d_{15}$ within the range 0.5 $d_{33} \le d_{15} \le$ 1.5$d_{33}$.

The fitting for strongly anisotropic domains, $r_s/l<<1$, is shown to lead to unphysically large tip sizes, $d\approx300$nm, or lateral domain shifts, $a\approx100$nm, inconsistent with $d\leq35$-40nm as determined from observed ferroelectric domain wall width.[21] To account for this discrepancy, we treated domain penetration length as a fitting parameter, and deconvolution results are shown in Fig. 2 (c-d). The fitting of piezoresponse loop and contrast of PFM images at small voltages $V$ suggest that the domains are oblate at nucleation and initial growth stages. This domain geometry necessitates efficient depolarization field screening in bulk of the sample, since in the rigid dielectric material depolarization field favors needle-like domain (depolarization energy is $\sim1/l^2$)[22]. This behavior is possible when screening charges are captured by the moving domain wall. For $d$=40nm, maximal domain radius $r_s$=300nm~10$d$ at bias 30V can be explained by effective mechanism of depolarization field surface screening in examined PZT film.

The careful inspection of domain evolution in Fig. 1 illustrates that often domain growth proceeds through the rapid jumps of domain walls, resulting in irregular domain shapes. Interestingly, these jumps can be associated with the fine structures of the local hysteresis loop, as illustrated in Fig. 3. Almost all hysteretic curves taken at higher magnification demonstrate visible kinks under increasing voltages. These kinks are reproducible upon repeating voltage excursions and thus could be attributed to the defects and associated pinning of domain walls. These features can be naturally explained by the presence of local defects[23] and long range electroelastic fields as it was theoretically predicted in the past (see, e.g., Ref. [24]). Based on result shown in Fig. 3, each reproducible feature on the hysteresis curve could be related to a separate metastable domain state pinned by the nearest defect site. Future improvement of the PFS technique is required to relate each kink in the hysteresis curve to an individual jump of the domain wall (nanoscale Barkhausen jump).

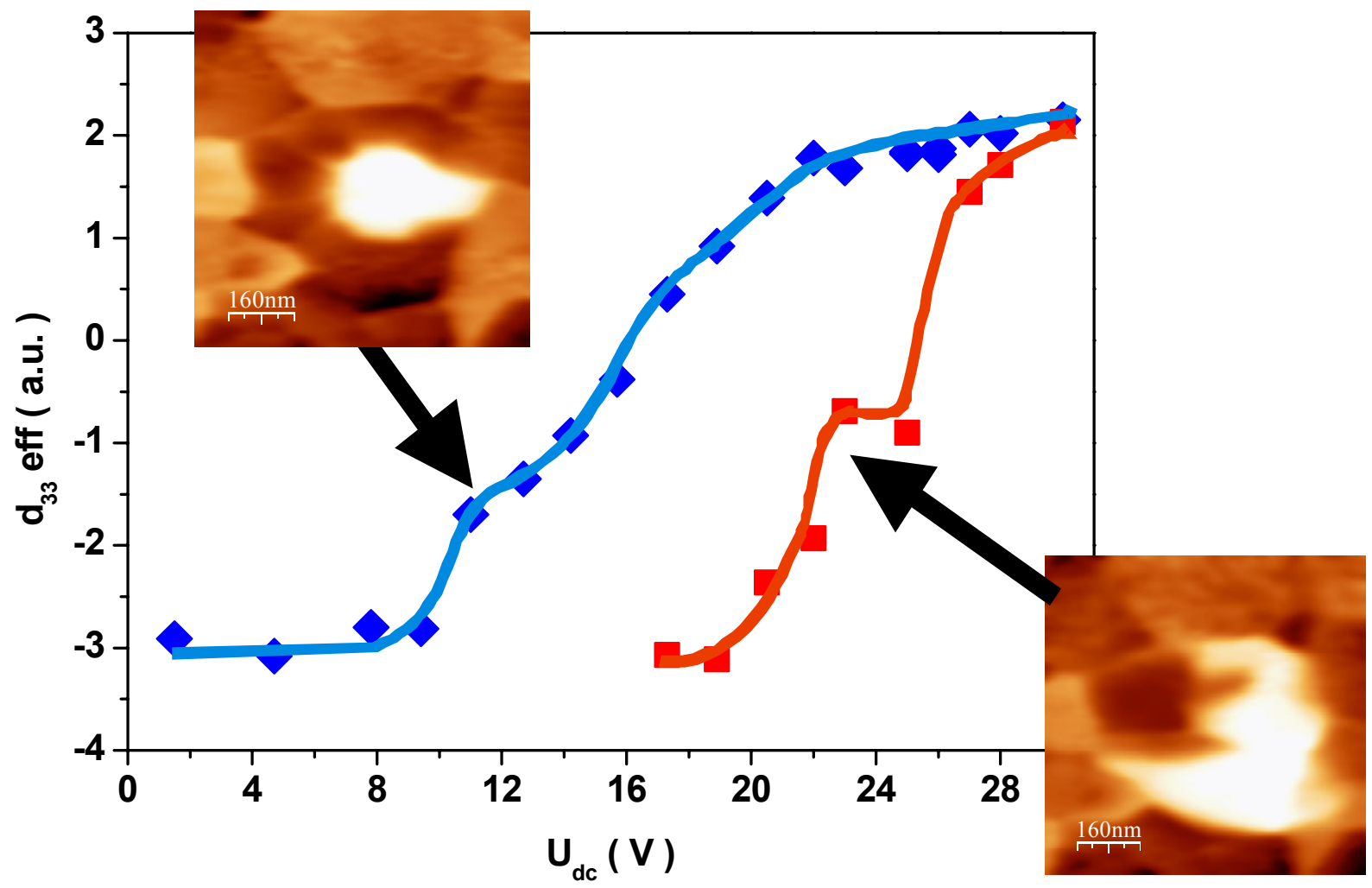

**Fig. 3.** Example of the proposed relationship between the fine structure of the piezoresponse hysteresis loops and jumps of domain walls observed in the investigated PZT films under increasing bias.

To summarize, we have analyzed hysteresis loop formation in Piezoresponse Force Microscopy. Direct comparison of the induced domain size and the PFS signal deconvoluted using self-consistent probe calibration has demonstrated that domain shape deviates significantly from simple needle-like geometry. In fact, the oblate domains are formed at nucleation and early growth stages. As a consequence, a new invariant $r_S^{9/2}/l$ was introduced for the description of the domain growth in polycrystalline PZT films. The fine structure features on piezoresponse spectra were demonstrated and tentatively related to the nanoscale jumps in domain geometry due to domain-defect interactions. Thus, high-resolution PFM spectroscopy offers the pathway to study the nanoscale mechanism of defect-mediated domain nucleation and growth.

Authors gratefully acknowledge multiple discussions and technical support of Dr. E. A. Eliseev. Research was sponsored in part (SVK) by the Division of Materials Sciences and Engineering, Office of Basic Energy Sciences, U.S. Department of Energy, under contract DE-AC05-00OR22725 with Oak Ridge National Laboratory, managed and operated by UT-Battelle, LLC. User proposal CNMS2007-265 (ORNL), FCT project PTDC/FIS/81442/2006 and FAME Nework of Excellence (NMP3-CT-2004-500159) are also acknowledged.

**Supplementary**

## 1) Appendix A

Measured in a PFS experiment is the electromechanical response related to the size of ferroelectric domain formed below the tip. Hence, to calculate the shape of the PFM hysteresis loop, the electromechanical response change induced by the domain is required. Within the framework of linearized theory by Felten et al,[17] the surface displacement vector $u_i(\mathbf{x})$ at position $\mathbf{x}$ is

$$u_i(\mathbf{x}) = \int_0^{\infty} d\xi_3 \int_{-\infty}^{\infty} d\xi_2 \int_{-\infty}^{\infty} d\xi_1 \frac{\partial G_{ij}(\mathbf{x},\xi)}{\partial \xi_l} E_k^p(\xi) d_{kmn}(\xi) c_{nmjl} \qquad \text{(A.1)}$$

where $\xi$ is the coordinate system related to the material, $G_{ij}(\mathbf{x},\xi)$ is Green tensor in isotropic elastic approximation [25] (anisotropy of elastic properties is assumed to be much smaller then that of dielectric and piezoelectric properties), $d_{nmp}$ are strain piezoelectric coefficients distribution, $c_{nmjl}$ are elastic stiffness and the Einstein summation convention is used. The electric field $E_k^p(\rho, x_3) = -\partial\varphi_p(\rho, x_3)/\partial x_k$ is created by the biased probe. Using an effective point charge approximation, the probe is represented by a single charge $Q$ located at distance, $d$, from a sample surface[13]. The potential $\varphi_p$ at $x_3 \geq 0$ has the form:

$$\varphi_p(\rho, x_3) \approx \frac{V d}{\sqrt{\rho^2 + (x_3/\gamma + d)^2}}. \qquad \text{(A.2)}$$

Here $\sqrt{x_1^2 + x_2^2} = \rho$ and $x_3$ is the radial and vertical coordinate respectively, $V$ is the bias applied to the probe, $\gamma = \sqrt{\varepsilon_{33}/\varepsilon_{11}}$ is the dielectric anisotropy factor; $d = 2R_0/\pi$ for a flattened tip represented by a disk in contact.

Integration of Eq. (A.1) for **x** = 0 yields the expression for effective vertical piezoresponse, $d_{33}^{eff}=u_3/V$, as[13]

$$d_{33}^{eff}(V)=d_{31}(f_1-2w_1(V))+d_{15}(f_2-2w_2(V))+d_{33}(f_3-2w_3(V)), \tag{A.3}$$

The functions $f_1=(2(1+\gamma)\nu+1)/(1+\gamma)^2$, $f_2=-\gamma^2/(1+\gamma)^2$, $f_3=-(1+2\gamma)/(1+\gamma)^2$ define the electromechanical response in the initial and final states of switching process;[26] $\nu$ is the Poisson ratio. The bias dependence of PFS signal, $w_i$, is determined by the domain sizes and shape. Domain shape affected by defects is approximated by cross-section with radius vector $r(\alpha)$ and length $l$ [see Fig.2(a)]. For this case, $w_i$ components are [26]:

$$w_1[r(\alpha),l]=\frac{1}{2\pi}\int_0^{2\pi}d\alpha\int_0^{\pi/2}d\theta\left(3\cos^2\theta-2(1+\nu)\right)\cos\theta\frac{R_S(\theta,r(\alpha),l)}{R_G(\theta,r(\alpha),l)}, \tag{A.4a}$$

$$w_2[r(\alpha),l]=\frac{3}{2\pi}\int_0^{2\pi}d\alpha\int_0^{\pi/2}d\theta\left(\frac{\gamma d+\cot\theta R_S(\theta,r(\alpha),l)}{R_G(\theta,r(\alpha),l)}-1\right)\cos^2\theta\sin\theta, \tag{A.4b}$$

$$w_3[r(\alpha),l]=-\frac{3}{2\pi}\int_0^{2\pi}d\alpha\int_0^{\pi/2}d\theta\cos^3\theta\frac{R_S(\theta,r(\alpha),l)}{R_G(\theta,r(\alpha),l)}. \tag{A.4c}$$

where the shape factors are introduced as $R_S(\theta,r,l)=rl\tan\theta/\sqrt{r^2+l^2\tan^2\theta}$ and $R_G(\theta,r,l)=\sqrt{(\gamma d+\cot\theta R_S(\theta,r,l))^2+\gamma^2R_S^2(\theta,r,l)}$. Using Lagrange mean point theorem, and Pade approximations theory, Eqs.(A.4) can be simplified as:

$$w_1[r(\alpha),l]\approx\frac{1}{2\pi}\int_0^{\pi/2}d\theta\frac{\left(3\cos^2\theta-2(1+\nu)\right)\cos\theta}{\sqrt{(\gamma d+\cot\theta R_{S1})^2+\gamma^2R_{S1}^2}}R_{S1}[r(\alpha)], \tag{A.5a}$$

$$w_2[r(\alpha),l]\approx\frac{3}{2\pi}\int_0^{\pi/2}d\theta\cos^2\theta\sin\theta\left(\frac{\gamma d+\cot\theta R_{S2}[r(\alpha)]}{\sqrt{(\gamma d+\cot\theta R_{S1})^2+\gamma^2R_{S1}^2}}-1\right), \tag{A.5b}$$

$$w_3[r(\alpha),l] \approx -\frac{3}{2\pi}\int_0^{\pi/2} \frac{d\theta \cos^3\theta \cdot R_{S3}[r(\alpha)]}{\sqrt{(\gamma d + \cot\theta R_{S3})^2 + \gamma^2 R_{S3}^2}}. \qquad \text{(A.5c)}$$

Where

$$R_{Si}[r(\alpha)] \approx \int_0^{2\pi} \frac{d\alpha \cdot rl \tan\theta_i^*}{\sqrt{r^2 + l^2 \tan^2\theta_i^*}} \qquad \text{(A.5d)}$$

and $\theta_i^*$ is mean point. Further interpolation by Pade at $a$=0, $\gamma \approx 1$, $r(\alpha) \approx$const and arbitrary $l$ (or alternatively $r \ll l$ and arbitrary $r(\alpha)$) leads to

$$R_{Si} \approx \frac{(\pi/6) l \cdot r_S}{\sqrt{r_S^2 + l^2(\pi/6)^2}} \qquad \text{(A.5e)}$$

for $i$=1,3, where

$$r_S \approx \frac{1}{2\pi}\int_0^{2\pi} d\alpha \cdot r(\alpha). \qquad \text{(A.5f)}$$

This approximation is good (less than 1-5% discrepancy) for $d_{33}$ contribution at arbitrary $r/l$ ratio, moderate for $d_{31}$ (about 25% discrepancy) and poor (more than 50% discrepancy) for $d_{51}$ for $r \gg l$, while $d_{51}$ contribution is negligibly small at $r \gg l$. Then direct integration on polar angle θ leads to expression:

$$d_{33}^{eff}(R_S(V)) \approx \left(\frac{3}{4}d_{33} + (1+4\nu)\frac{d_{31}}{4}\right)\frac{\pi d - 8R_S(V)}{\pi d + 8R_S(V)} + \frac{d_{15}}{4}\frac{3\pi d - 8R_S(V)}{3\pi d + 8R_S(V)} \qquad \text{(A.6a)}$$

Under the condition $0 < a \ll r_S$ and $r_S \ll l$ one obtains using concept of effective piezoresponse volume that

$$d_{33}^{eff}(r_S, a) \approx d_{33}^{eff}(r_S - a) + \left(d_{33}^{eff}(r_S + a) - d_{33}^{eff}(r_S - a)\right)\frac{2r_S - a}{4r_S}, \qquad \text{(A.6b)}$$

and therefore:

$$d_{33}^{eff}(r_S,a) \approx \left( \begin{array}{l} \frac{3}{4} d_{33}^{*} \left( 1 - \frac{16 r_S}{\pi d + 8 r_S} + 8\pi d \frac{(\pi d + 24 r_S) a^2}{r_S (\pi d + 8 r_S)^3} \right) + \\ + \frac{d_{15}}{4} \left( 1 - \frac{16 r_S}{3\pi d + 8 r_S} + 24\pi d \frac{(3\pi d + 24 r_S) a^2}{r_S (3\pi d + 8 r_S)^3} \right) \end{array} \right) \qquad (A.7)$$

The case $2a \sim r_S$ should fitted numerically on the base of exact expressions (A.4).

## Appendix B

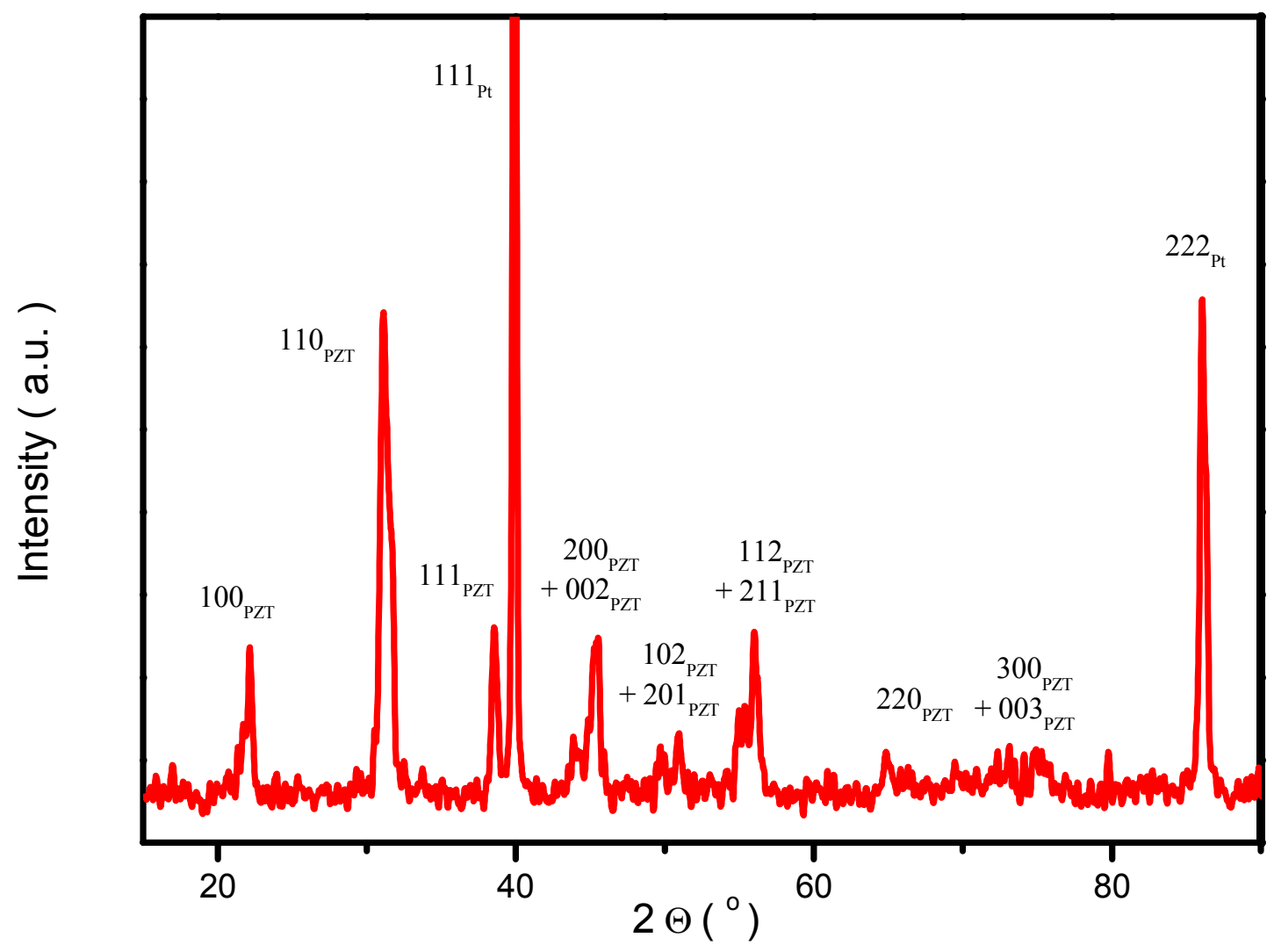

Fig. X-ray diffraction pattern of $Pb(Zr_{1-x}Ti_x)O_3$, x = 0.7, thickness 4 µm.

This is no textured film. All orientations are present in the film.

Data on this film **$Pb(Zr_{1-x}Ti_x)O_3$ x = 0.7** (INOSTEK):

| | |
|---|---|
| Thickness | 4 µm |
| Orientation | Random (XRD plots available) |
| Dielectric constant | ~450 (poled) |
| | ~600 (unpoled) |
| Remanent polarization | 35 μC/cm$^2$ |
| Coercive field | 75 kV/cm |
| $d_{33}$ (clamped) | 50 pm/V |
| $e_{31}$ (wafer texture technique) | -2 C/m2 |

Assuming that values of piezoelectric modulus from the table are effective values, related to intrinsic ones as $d_{33}^{film} = d_{33} - d_{31} \frac{2s_{13}^E}{s_{11}^E + s_{12}^E}$, $e_{31}^{film} = e_{31} - e_{33} \frac{c_{13}^E}{c_{33}^E}$ (see e.g. Ref. [27]) and using elastic compliances from Ref. [[28]] $s_{11}$=8.5 and $s_{12}$=-2.8 $10^{-12}$ Pa$^{-1}$, we obtained $d_{31}$=-11.4, and $d_{33}$=61.2 pm/V. The deconvolution was performed for $d_{15}$=0.5$d_{33}$ for bulk and $d_{15}$=1.5$d_{33}$ for ceramics, and the results are shown to be weakly dependent on $d_{15}$.